\documentclass[aps,twocolumn,prd,amsmath,footinbib,preprintnumbers,superscriptaddress]{revtex4-1}
\usepackage{epsfig,slashed,nicefrac}
\usepackage[utf8]{inputenc}
\usepackage{color}
\usepackage{multirow}
\usepackage{float} 
\usepackage{subfig} 
\usepackage{eqnarray,amsmath}
\usepackage{paralist}
\usepackage{times}
\usepackage{txfonts}
\usepackage{comment}
\usepackage[normalem]{ulem}

\newcommand{\eg}{{\it e.g.}}
\newcommand{\ie}{{\it i.e.}}
\newcommand{\cf}[1]{{Fig.~\ref{#1}}}

\def\RpA     {\mbox{$R_{pA}$}}

\def\RdAu    {\mbox{$R_{d\rm Au}$}}
\def\RpPb    {\mbox{$R_{p\rm Pb}$}}
\def\RgPb    {\mbox{$R_g^{\rm Pb}$}}
\def\pPb  {$p\mathrm{Pb}$}

\def\be{\begin{equation*}}
\def\ee{\end{equation*}}
\def\bsp#1\esp{\begin{split}#1\end{split}} 
\def\bpm{\begin{pmatrix}}
\def\epm{\end{pmatrix}}

 \usepackage{ifpdf}
 \ifpdf
 \usepackage[pdftex]{hyperref}
 \else
 \usepackage[hypertex]{hyperref}
 \fi

 \hypersetup{
   pdftitle={},%
   pdfauthor={},%
   pdfsubject={},%
   pdfkeywords={},%
   pdfstartview={},%
   bookmarksopen=true, breaklinks=true, debug=true, %
   colorlinks=true, linkcolor=red, citecolor=blue, urlcolor=blue
 }

\begin{document}

\title{Gluon Shadowing in Heavy-Flavor Production at the LHC}

\author{Aleksander Kusina}
\affiliation{Institute of Nuclear Physics, Polish Academy of Sciences, ul. Radzikowskiego 152, 31-342 Cracow, Poland}

\author{Jean-Philippe Lansberg}
\affiliation{IPNO, CNRS-IN2P3, Univ. Paris-Sud, Universit\'e Paris-Saclay, 91406 Orsay Cedex, France}

\author{Ingo Schienbein}
\affiliation{Laboratoire de Physique Subatomique et de Cosmologie,
              Universit\'e Grenoble-Alpes, CNRS/IN2P3,
              53 avenue des Martyrs, 38026 Grenoble, France}

\author{Hua-Sheng Shao}
\affiliation{Sorbonne Universit\'es, UPMC Univ. Paris 06, UMR 7589, LPTHE, F-75005, Paris, France}
\affiliation{CNRS, UMR 7589, LPTHE, F-75005, Paris, France}

\date{\today}

\begin{abstract}
We study the relevance of experimental data on heavy-flavor 
[$D^0$, $J/\psi$, $B\rightarrow J/\psi$ and $\Upsilon(1S)$ mesons]
production in proton-lead collisions at the LHC  
to improve our knowledge of the gluon-momentum distribution inside heavy nuclei.
We observe that the nuclear effects encoded in both most recent global fits of 
nuclear parton densities at next-to-leading order (nCTEQ15 and EPPS16) provide
 a good overall description of the LHC data. We interpret this 
as a hint that these are the dominant ones.
In turn, we perform a Bayesian-reweighting analysis for each particle data sample
which shows that each of the existing heavy-quark(onium) data set clearly points --with a minimal statistical 
significance of 7 $\sigma$-- to a 
shadowed gluon distribution at small $x$ in the lead. 
Moreover, our analysis 
corroborates the existence of gluon antishadowing.
Overall, the inclusion of such heavy-flavor data in a global fit would 
significantly reduce the uncertainty on the gluon density down to  
$x\simeq 7\times 10^{-6}$ --where no other data exist-- while keeping an 
agreement with the other data of the global fits. Our 
study accounts for the factorization-scale uncertainties which dominate
for the charm(onium) sector.
\end{abstract}

\maketitle

\textit{Introduction} -- Parton-distribution functions (PDFs), describing the longitudinal-momentum distributions
of quarks and gluons inside hadrons, provide the essential link between the measurable
hadronic cross sections and the perturbatively-calculable cross sections of high-energy
processes induced by quarks and gluons. The precise determination of PDFs of {\it protons}, $f_i^p$, is an extremely active area of research where several groups perform global
analyses of a wide variety of experimental hard-process data. The modern global analyses 
\cite{Dulat:2015mca,Harland-Lang:2014zoa,Ball:2017nwa,Accardi:2016qay,Alekhin:2017kpj,Alekhin:2014irh}
have evolved into impressive ventures with state-of-the-art perturbative calculations and sophisticated statistical
methods to extract optimum PDFs along with their uncertainties. 

The situation is more challenging --but not less interesting-- for PDFs of {\it nucleons inside nuclei}, $f_i^{[p,n]/A}$,
with nuclear data significantly more complex to collect and with two additional degrees of freedom,
the number of protons ($Z$) and neutrons ($N=A-Z$) in the studied
nuclei. Nuclear PDFs (nPDFs) are key ingredients to use perturbative probes of the quark-gluon
plasma produced in ultrarelativistic nucleus-nucleus collisions at RHIC and the LHC~\cite{Andronic:2015wma}.
As such, their determination goes even beyond the understanding of the nucleus content in terms of quarks and gluons.
Since the early 1980s, we know that the nuclei are not a simple collection of free nucleons,
and nPDFs are not equal to a sum of nucleon PDFs.
In fact, the corresponding analyses
rather bear on nuclear-modification factors (NMF), like in lepton-nucleus ($\ell A$) collisions
$R[F_2^{\ell A}]= F_2^{\ell A}/(Z F_2^{\ell p}+(A-Z) F_2^{\ell n})$ 
for the deep-inelastic scattering (DIS) structure function $F_2$
and parton-level NMFs $R_i^A(x,\mu_F) = f_i^A/(Z f_i^p + N f_i^n)$ with 
$f_i^A \equiv Z f_i^{p/A} + N f_i^{n/A}$ ($i=g, q, \bar q$), instead of the absolute nPDFs.

Based on earlier studies of $F_2$ \cite{Aubert:1983xm,Goodman:1981hc,Bodek:1983ec,Bari:1985ga,Benvenuti:1987az,Ashman:1988bf,Arneodo:1988aa}, one knows that, for the {\it quarks},
\begin{inparaenum}[(i)]
\item $R_q^A>1$ for $x\gtrsim0.8$ (Fermi-motion region), 
\item $R_q^A<1$ for $0.25\lesssim x\lesssim0.8$ (EMC region), 
\item $R_q^A>1$ for $0.1\lesssim x\lesssim0.25$ (antishadowing region), and 
\item $R_q^A<1$ for $x\lesssim0.1$ (shadowing region) 
\end{inparaenum}
where different physics mechanisms were proposed to explain
this behavior.
At medium and large longitudinal-momentum fractions, $x$, $R_q^A$ is usually
explained by nuclear-binding and medium effects and the Fermi motion
of the nucleons \citep{Geesaman:1995yd} but a fully conclusive picture
has not yet emerged after the discovery of the  EMC effect~\cite{Higinbotham:2013hta}.
At small $x$, coherent scatterings inside the nucleus 
 explain the observed suppression of $F_2$, referred to as shadowing.
Antishadowing is even less understood. 
Therefore, just like in the nucleon case, nPDFs are determined by performing global analyses 
of experimental data \cite{Eskola:2016oht,Kovarik:2015cma,Khanpour:2016pph,deFlorian:2011fp,Hirai:2007sx}.

Compared to the {\it quark} content --directly probed by data on $\ell A$ DIS and the proton-nucleus ($pA$) Drell-Yan  process--, 
the {\it gluon} content of the nuclei is even less known. 
To compensate this lack of constraints, both most recent global next-to-leading order (NLO) analyses of nPDFs,
nCTEQ15~\cite{Kovarik:2015cma} and EPPS16~\cite{Eskola:2016oht},
used RHIC pion and LHC jet data (in case of EPPS16) to constrain the gluon densities
down to $x\sim 10^{-3}$. However, there is no data at $x \lesssim 10^{-3}$.
Hence, we do not know anything about the gluon at small $x$; 
the gluon nPDFs in this region are obtained by extrapolating nPDFs from larger $x$ region. As such, they
essentially depend on the parametrizations of the $x$-dependence 
of nPDFs at the initial scale $\mu_{F,0} \sim 1$~GeV. 

As discussed in Refs.~\cite{Stavreva:2010mw,Helenius:2016hcu}, this lack of knowledge of 
the gluon nPDF is thus {\it a priori} not reflected by the set of error PDFs  
provided together with the best fit PDFs. Accordingly, increasing the flexibility of
the initial nPDF parametrizations leads to much larger uncertainties in this region
as evidenced by the EPPS16 set as opposed to the EPS09~\cite{Eskola:2009uj} and nCTEQ15 ones.
Clearly, a determination of the small-$x$ gluon nPDFs and the reduction of their uncertainties is necessary for the heavy-ion phenomenology.

Recently, using heavy-flavor (HF) production at the LHC was proposed 
for an improved determination of the small-$x$ gluons in the {\it proton}~\cite{Zenaiev:2015rfa,Gauld:2015yia,Cacciari:2015fta,Gauld:2016kpd,deOliveira:2017ega}. We also noticed an earlier proposal~\cite{Gauld:2015lxa}.
Motivated by the results of these studies, we 
performed the first analysis of the impact of heavy-quark(onium) data
in LHC proton-lead ($p$Pb) collisions on the determination of 
nPDFs (nCTEQ15, EPPS16) as a way to constrain the small-$x$ gluon density 
in lead down to $x \simeq 7 \times 10^{-6}$.

The interpretation of our results depends on the
reliability of nPDF factorization in the nuclear environment, which is a question of considerable
theoretical and practical importance.
In this context, we note that other cold-nuclear matter (CNM) effects~\cite{Gerschel:1988wn,Vogt:1999cu,Ferreiro:2014bia,Capella:2005cn,Capella:2000zp,Gavin:1990gm,Arleo:2012hn,Sharma:2012dy,Arleo:2010rb,Brodsky:1992nq,Gavin:1991qk,Brodsky:1989ex,Ducloue:2015gfa,Ma:2015sia,Fujii:2013gxa,Qiu:2013qka,Kopeliovich:2001ee,Ferreiro:2008wc,Ferreiro:2011xy,Ferreiro:2013pua,Vogt:2010aa}
could become relevant in some specific conditions, in particular for the quarkonium case.
In our study, they can be seen as higher-twist (HT) contributions and
the use of leading-twist (LT) factorization becomes a working assumption to be tested. 
Once validated by data, as we will show, this assumption of LT factorization can be
employed to learn about the internal structure of the nucleus.

\textit{Methodology} -- 
The cross sections measured in $pA$ collisions at colliders are nearly always
normalized to the $pp$ ones~\cite{Albacete:2016veq,Andronic:2015wma,Albacete:2017qng} since one is primarily interested in deviations from the free nucleon case, up to isospin effects. 
For DIS off a nucleus $A$, and thus $F_2^{\ell A}$, the NMF $R[F_2]$ is directly related
to the modification $R_q^A$ of the (anti)quark nPDF compared to its PDF.
For the gluons, one similarly defines
$R_g^A$ entering theoretical evaluations of the NMF $\RpA\equiv{d\sigma_{pA}}/{(A \times d\sigma_{pp})}$,
which can be differential in the transverse momentum ($P_{T,\mathcal{H}}$) or the center-of-momentum (c.m.s.)
rapidity $y_{{\rm c.m.s.},\mathcal{H}}$ of the hadron $\mathcal{H}$. 
The nPDF sets provide parametrization of $R_g^A$ at any $x$ and
scale. In the absence of nuclear effects, 
$R_g^A=1$ and we observe $\RpA(O_{\mathcal{H}})\simeq1$. 
Unlike the simple case of $F_2$ at leading order, $R^A_g$ enters $\RpA$ 
via a convolution which requires
a control of the parton-scattering kinematics.

The focus on $\RpA$ has several advantages. It allows to leave aside, in the theory 
evaluation, the {\it proton} PDF uncertainty at very small $x$ which may not 
always be negligible. 
Second, $\RpA$ is in general less sensitive to QCD corrections 
which may affect the normalization of the 
cross-section predictions. Third, some experimental uncertainties cancel 
in $\RpA$ and, at the LHC, \RpPb\ is usually more precise than the corresponding 
\pPb\ cross sections.

To connect $\RpA$ and $R_g^A$, we will use the data-driven approach~\cite{Lansberg:2016deg,Shao:2012iz,Shao:2015vga} where the parton-scattering-matrix 
 elements squared $|A|^2$ are parametrized into empirical functions and determined from $pp$ data assuming a $2\to2$ 
kinematics together with given proton PDF, where we choose CT14NLO~\cite{Dulat:2015mca}. It was first motivated to bypass the complications inherent 
to our lack of understanding of the quarkonium-production mechanisms
(see \eg, Refs.~\cite{Andronic:2015wma,Brambilla:2010cs}) whereas it suffices to
evaluate the nPDF effects in $\RpA$. Such an approach also applies to open
HF hadrons~\cite{Lansberg:2016deg}. In the latter cases, full-fledged perturbative QCD 
computations exist~\cite{Kniehl:2004fy,Kniehl:2005mk,Kniehl:2012ti,Kniehl:2011bk,Kniehl:2015fla,Alwall:2014hca,
Cacciari:1998it,Cacciari:2001td,Cacciari:2012ny,Mangano:1991jk,Nason:2004rx,Frixione:2007vw,Frixione:2002ik,Frixione:2003ei}
which we have used to validate the method.
As in Ref.~\cite{Lansberg:2016deg}, we use a specific functional form for $|A|^2$
proposed in Ref.~\cite{Kom:2011bd} to model single-quarkonium hadroproduction for double-parton scattering~\cite{Kom:2011bd,Lansberg:2014swa,Lansberg:2015lva,Shao:2016wor,Borschensky:2016nkv},
which is sufficiently flexible to give a good description
of single-inclusive-particle production.

There are several advantages in using this approach: 
\begin{inparaenum}[(i)]
\item the uncertainty in the $pp$ cross section is controlled by the measured data,
\item it can be applied to any single-inclusive-particle spectrum as long as 
the relative weights of the different channels (parton luminosities times $|A|^2$) are known, and
\item the event generation is much faster than with QCD-based codes, allowing us to study 
several nPDFs with several scale choices in an acceptable amount of computing time.
\end{inparaenum}
Indeed, to quantify the intrinsic theoretical uncertainty from the 
factorization scale $\mu_F$, we have varied it about a default scale $\mu_0$ as 
$\mu_F=\xi \mu_0$ with $\xi = 0.5,1.0,2.0$. $\mu_0^2$ was taken to be $M_{\cal Q}^2+P_{T,{\cal Q}}^2$
for ${\cal Q}=(J/\psi,\Upsilon)$, $4 M_{D}^2+P_{T,{D}}^2$ for $D^0$, and $4M_B^2 + ({M_B}/{M_{J/\psi}})^2 \times P_{T,J/\psi}^2$ for $B \to J/\psi$.

Compared to Ref.~\cite{Lansberg:2016deg}, the $pp$ baseline study was improved.
For the first time, we considered the $B \to J/\psi$ data. For $D^0$, $J/\psi$ 
and $\Upsilon(1S)$, we advanced the scale study with a variation in the $pp$
baseline itself and not only in $R_g^{\text{Pb}}(x,\mu_F)$, where $pp$ fits were done with each scale choice. As what concerns the \RpPb\ results, 
we checked that, for the cases of $D^0$ and 
$B\rightarrow J/\psi$ production, the scale uncertainty is nearly identical to that
with the ``fixed-order-plus-next-to-leading log" (FONLL)~\cite{Cacciari:1998it,Cacciari:2001td,Cacciari:2012ny} calculation
(see a comparison in the Supplemental Material).
As expected, FONLL gives much larger scale uncertainties on the {\it yields}.

As announced, to study the impact of HF experimental data  on the gluon nPDF determination without performing
a full fit, we employed the Bayesian-reweighting method~\cite{Giele:1998gw,Ball:2010gb,Ball:2011gg,Sato:2013ika,Paukkunen:2014zia,Kusina:2016fxy}. 
This method is a direct application of Bayes theorem allowing one to include new data into a given PDF analysis without a fit. For the present study, we followed the same approach as in 
Ref.~\cite{Kusina:2016fxy}. 
Since both nCTEQ15 and EPPS16 are Hessian nPDFs, we converted the
Hessian error PDFs into $10^4$ Monte Carlo replicas, representing the underlying 
probability distribution~\footnote{The 32 (40) eigensets of these Hessian nPDFs are meant to
display a 90\% confidence level (CL) uncertainty with a tolerance 
$\Delta \chi^2=35$ (nCTEQ15) and $52$ (EPPS16) respectively. Let us note that in what follows our results will be displayed at 68\% CL, \ie\ 1-$\sigma$ which is more suitable for theory-data comparison with 1-$\sigma$
experimental uncertainties. This simply amounts to reduce the uncertainties by $\sqrt{2}{\rm erf}^{-1}(0.90)\simeq 1.645$.}. For each PDF replica, one computes the $\chi^2$ of the considered data which is used to reweight them. 
Replicas describing better the data get larger weights than those unfavored 
by them. Hence, one obtains a modified probability distribution of the nPDFs 
like a fit would do.

\textit{Data selection} -- 
Like for all global PDF fits, a data selection is in order to avoid HT corrections. In our
case, it is also important to select a kinematical region where gluon fusion dominates and other effects are
negligible. As such, we considered the HF production in $pA$ collisions at LHC energies.
In the quarkonium case, due to the large Lorentz boost at these energies, 
the heavy-quark pair remains almost pointlike all along its way through the nuclear matter. 
Therefore, breakup~\cite{Vogt:2004dh,Lourenco:2008sk}, thought to be important 
at lower energies, is negligible at the LHC. 
We focused on $J/\psi$ and $\Upsilon(1S)$ to limit the contamination 
by possible comover effects~\cite{Gavin:1990gm,Capella:2000zp,Capella:2005cn,Ferreiro:2014bia,Ferreiro:2018wbd}, 
on the more fragile excited states [$\psi(2S)$, $\Upsilon(2S)$, $\Upsilon(3S)$]. 

Overall, this gives the  ALICE~\cite{Abelev:2014hha} and LHCb~\cite{Aaij:2017gcy} $D^0$ data;
the  ALICE~\cite{Adam:2015iga,Abelev:2013yxa} and LHCb~\cite{Aaij:2013zxa,Aaij:2017cqq} $J/\psi$ data;
the LHCb~\cite{Aaij:2017cqq} $B\rightarrow J/\psi$ data; the 
 ALICE~\cite{Abelev:2014oea}, ATLAS~\cite{TheATLAScollaboration:2015zdl}  and LHCb~\cite{Aaij:2014mza} $\Upsilon(1S)$ data.
We could also add the $d$Au $J/\psi$ RHIC data. Instead, we preferred to focus
on the LHC data at 5 and 8 TeV and to use the RHIC~\cite{Adare:2010fn,Adare:2012qf} and the new LHC~\cite{Aaboud:2017cif,Adam:2016ich} ones as cross checks.

\begin{figure*}[hbt!]
\centering
\subfloat[Prompt $D^0$]{\includegraphics[width=0.66\columnwidth,draft=false]{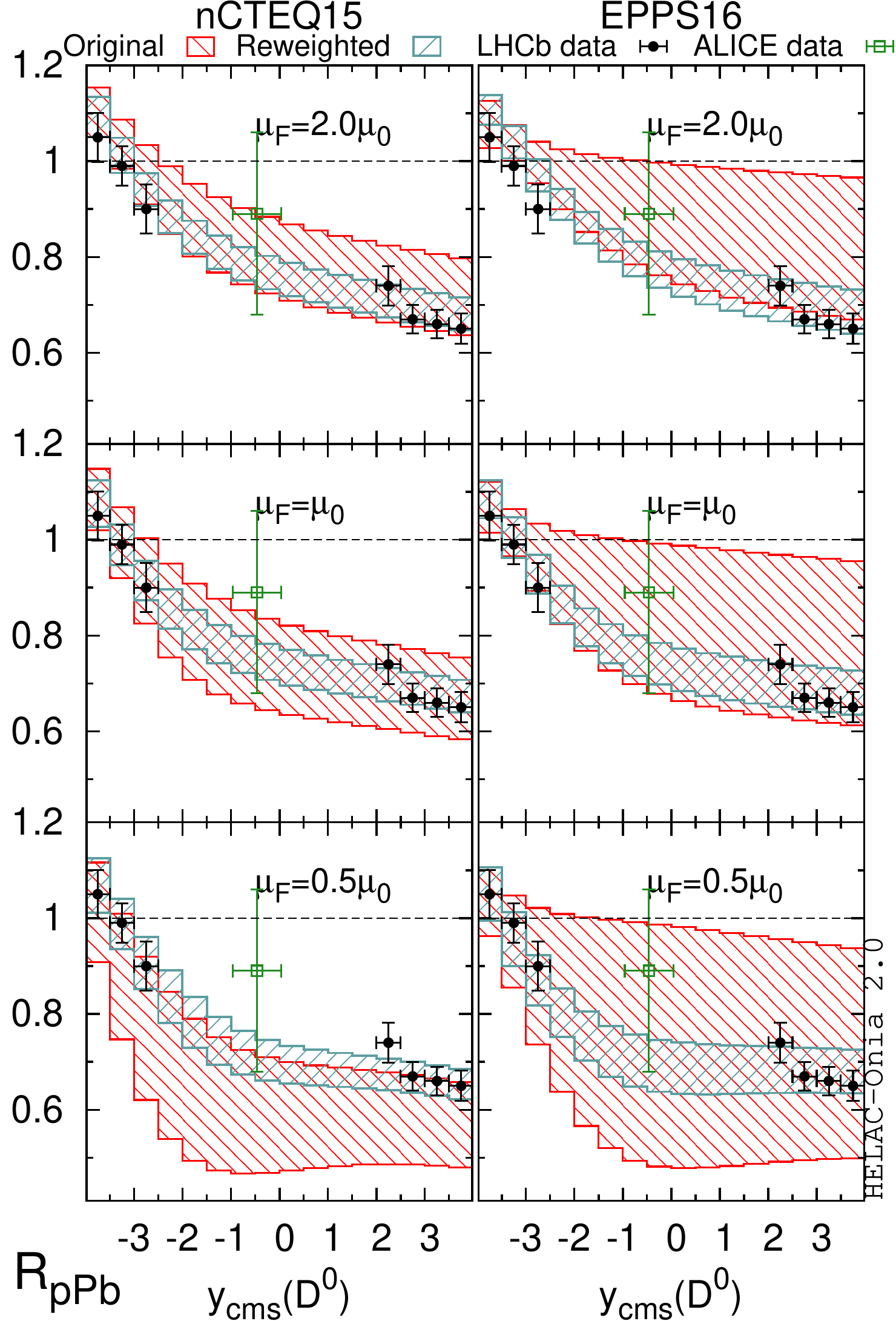}\label{diagram-a}}
\subfloat[Prompt $J/\psi$]{\includegraphics[width=0.66\columnwidth,draft=false]{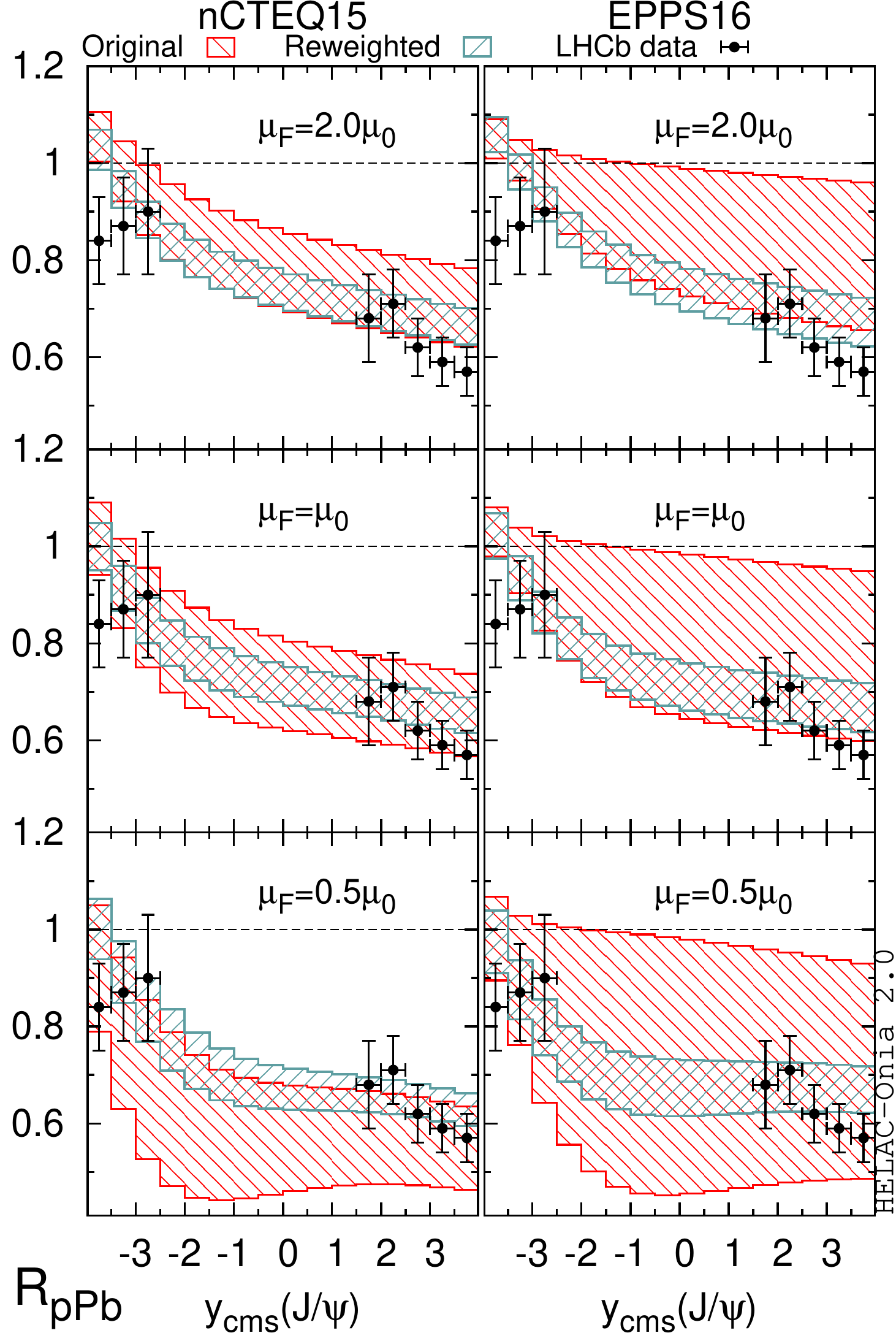}\label{diagram-b}}
\subfloat[$B\rightarrow J/\psi$]{\includegraphics[width=0.66\columnwidth,draft=false]{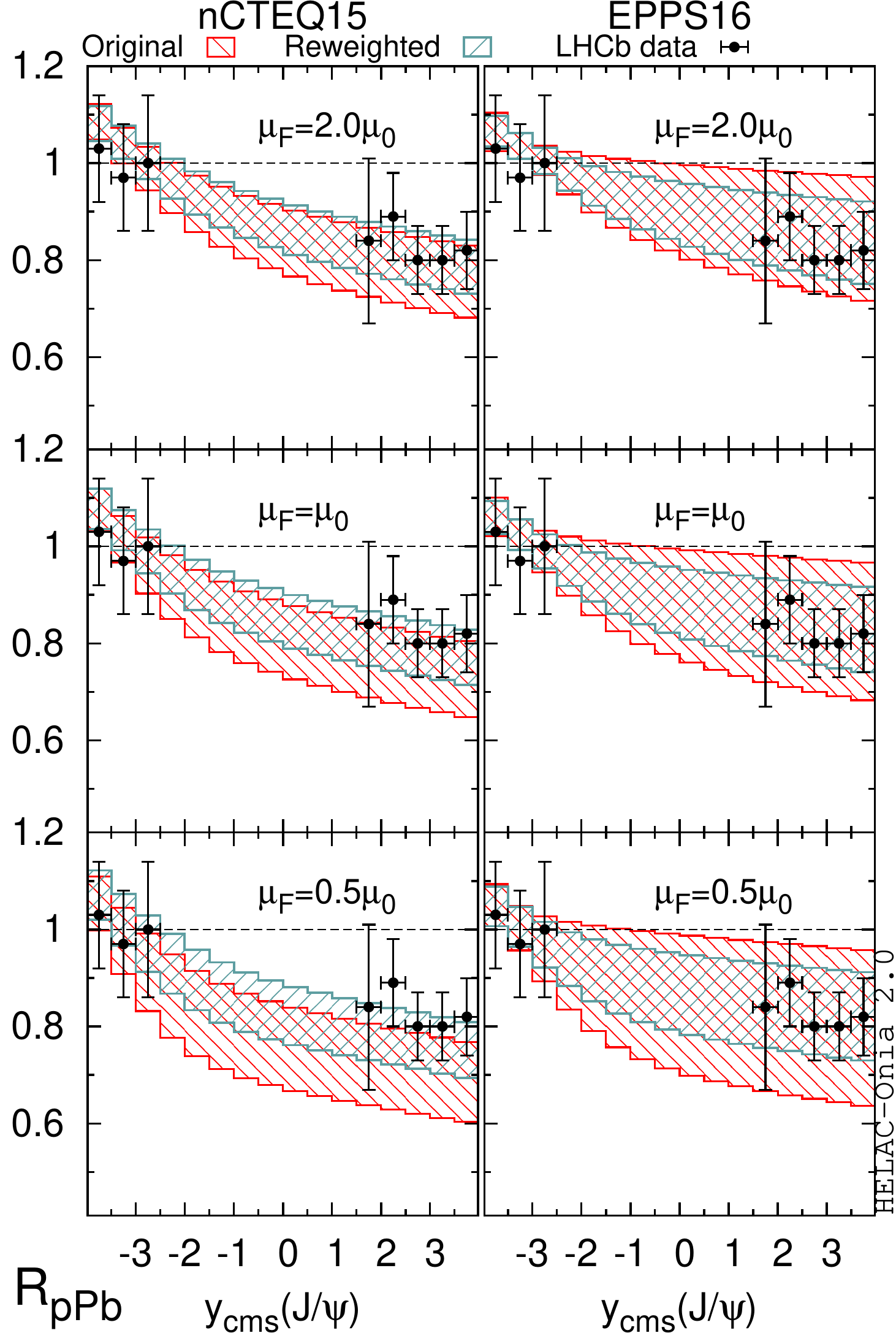}\label{diagram-c}}\\
\subfloat[$\Upsilon(1S)$]{\includegraphics[width=0.66\columnwidth,draft=false]{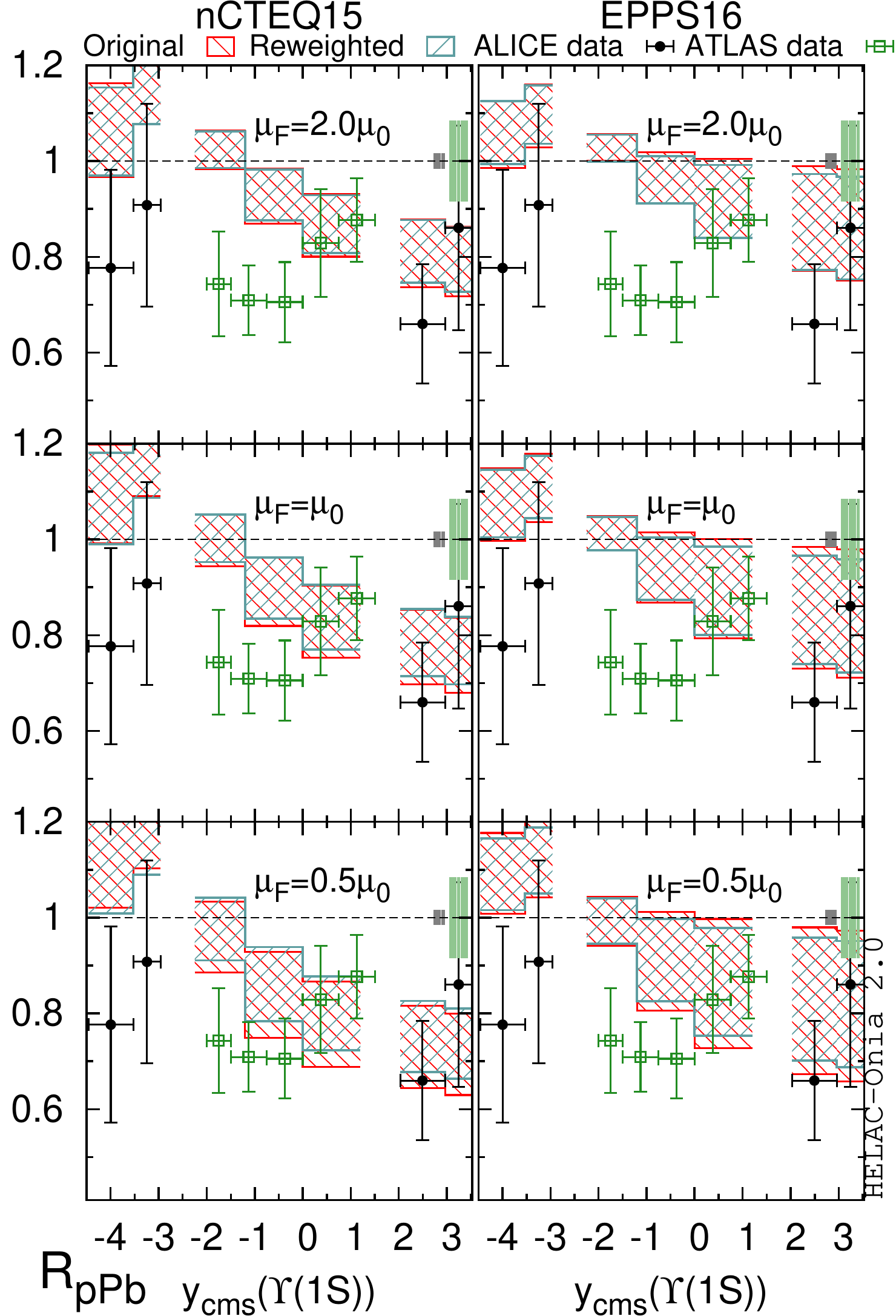}\label{diagram-d}}
\subfloat[nCTEQ15 nPDF]{\includegraphics[width=0.66\columnwidth,draft=false]{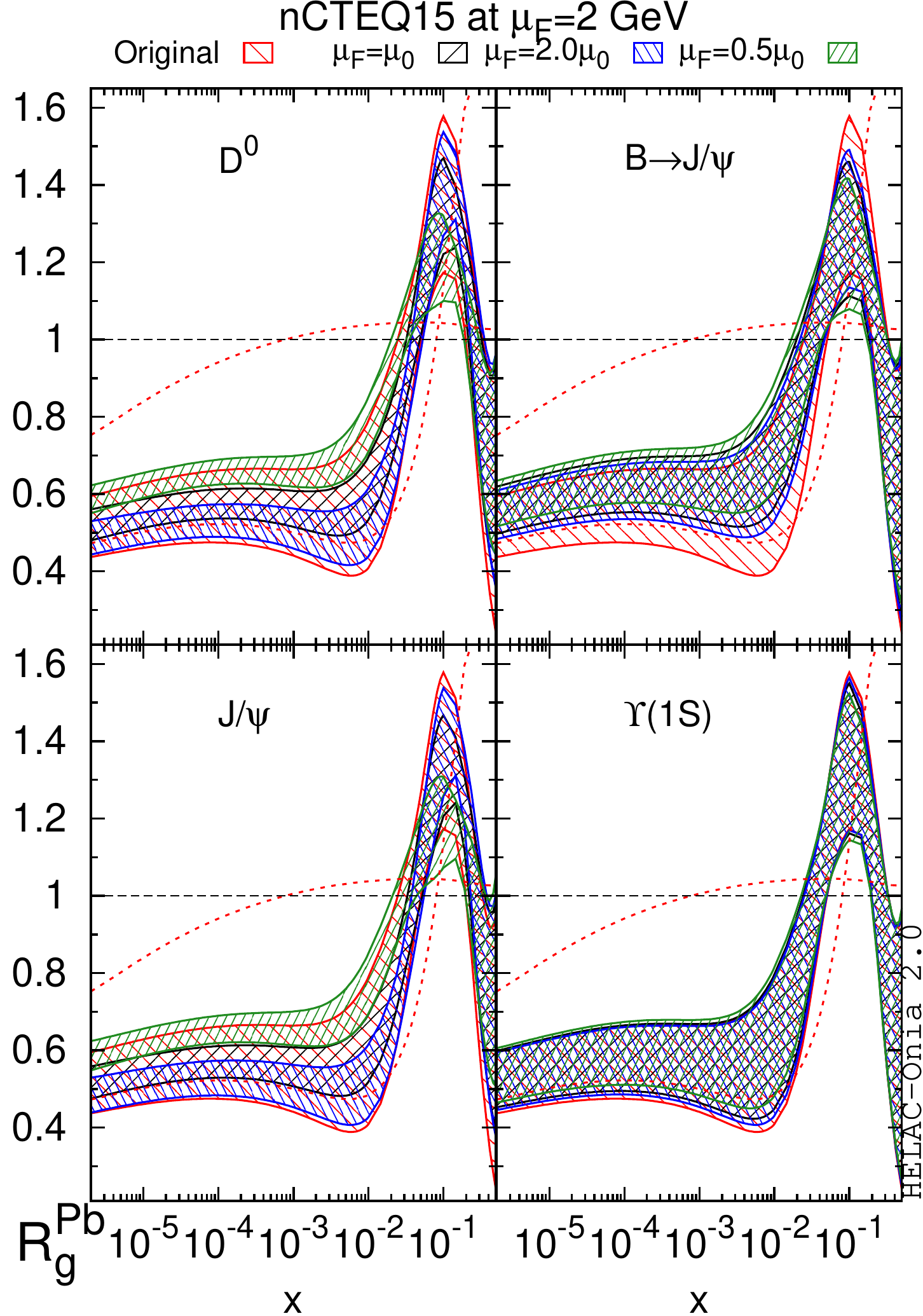}\label{diagram-e}}
\subfloat[EPPS16 nPDF]{\includegraphics[width=0.66\columnwidth,draft=false]{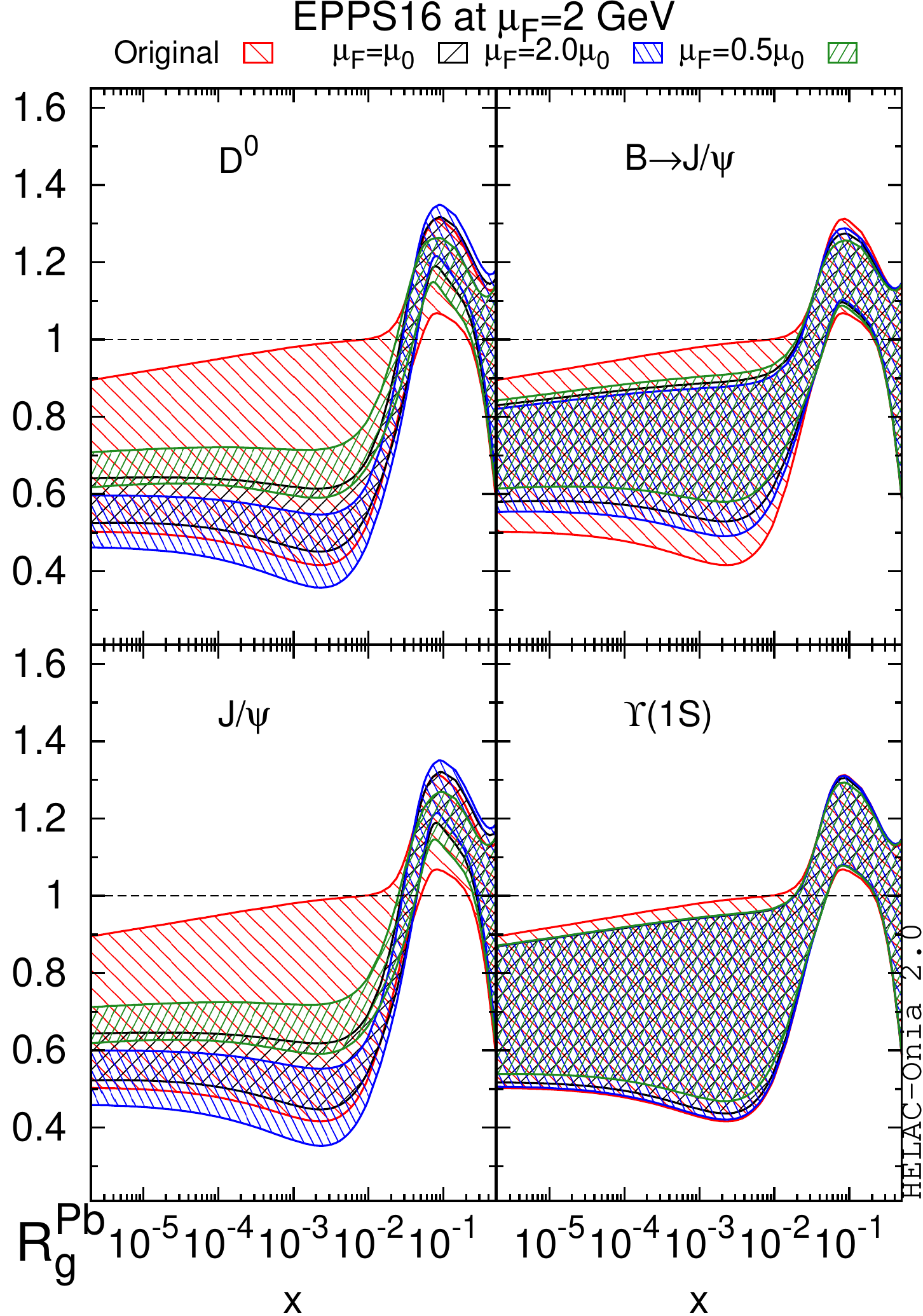}\label{diagram-f}}
\caption{Selected \RpPb\ results before and after reweighting for (a) prompt $D^0$, (b) prompt $J/\psi$, (c) $B\rightarrow J/\psi$, (d) $\Upsilon(1S)$ as well as the final reweighted nPDF uncertainties (e) nCTEQ15 and (f) EPPS16 with constraints from both $R_{p{\rm Pb}}$ vs $P_{T,\mathcal{H}}$ and $y_{{\rm c.m.s.},\mathcal{H}}$ data. The shown experimental data are from Refs.~\cite{Aaij:2017gcy,Adam:2016ich,Aaij:2017cqq,Abelev:2014oea,Aaboud:2017cif}. The error bands due to nPDF uncertainty are given at $68\%$ C.L.}
\label{diagrams} \vspace*{-0.5cm}
\end{figure*}

\textit{Results} --
Figs.~(\ref{diagram-a}--\ref{diagram-d}) show 
a representative comparison of our theoretical 
calculations with the data for $D^0$, $J/\psi$, $B\rightarrow J/\psi$ and $\Upsilon(1S)$. 
The NMF obtained with nCTEQ15 and EPPS16 have significantly different 
central values and uncertainties but both agree with the data.
This observation is striking as the used gluon nPDFs were derived from totally different
observables like DIS and Drell-Yan processes, and yet they allow us to reproduce the most important feature of the
data~\cite{Lansberg:2016deg} which makes our reweighting analysis meaningful.
We see this as a confirmation of the LT factorization
(see also Refs.~\cite{Frankfurt:2011cs,Frankfurt:2016qca,Guzey:2013xba,Abelev:2012ba}).

As for the reweighting results (gray-blue hatched bands in Figs.~[\ref{diagram-a}--\ref{diagram-d}]), if we could simply fix the scale to a single 
value for each particle, the LHC \RpPb\ data for prompt $D^0$ and $J/\psi$ would reduce 
the uncertainties of the gluon density by a factor 3 for EPPS16
and 2 for nCTEQ15 down to $x\simeq 7\times 10^{-6}$ [compare the gray-blue and red hatched 
bands in Figs.~\ref{diagram-a} and \ref{diagram-d}).  
The current $B\rightarrow J/\psi$ and $\Upsilon(1S)$ data 
do not constrain the gluon nPDFs due to their large uncertainties and relatively large scales. Yet, the larger 
samples collected at 8 TeV should improve the situation.

We now discuss the scale uncertainties and recall that
$d\sigma_{p\rm Pb} \sim  f_g^p (f_g^p \RgPb)\otimes |A|^2$.
Because of QCD evolution, a larger $\mu_F$  implies a \RgPb\ closer to unity
together with a smaller PDF uncertainty.
Indeed, the bands in Figs.~(\ref{diagram-a}--\ref{diagram-d}) are closer to unity and shrink from
$\mu_F = 0.5 \mu_0$ to $\mu_F = 2 \mu_0$. For nCTEQ15, such variations for $D^0$ and
$J/\psi$ are even similar to the nPDF uncertainty itself. 

Clearly, such a scale ambiguity should impact the reweighting results even though 
the (gray-blue) reweighted bands seems not to show such a sensitivity. It is 
perfectly normal since the replicas are to match the data. The key point is 
that they match it at different scales.
Consequently, when the reweighted bands are evolved to a common scale 
$\mu_F=2$~GeV, the reweighted nPDF uncertainties obtained with different 
scales do not superimpose (compare the black, blue and green bands in 
Figs.~(\ref{diagram-e}--\ref{diagram-f})).

The envelope of these scale-induced variations is about twice as large as their 
width for the $D^0$ and $J/\psi$ cases, confirming that the scale uncertainty 
must be accounted for to obtain reliable uncertainties from these precise
data. For the heavier bottom(onium) states, the scale uncertainty
is not only much smaller than the nPDF uncertainties but also very small in absolute
value, which implies that more precise data could play a major role
for a precision determination of the gluon nPDF at small $x$.

Despite these uncertainties, our results are striking: the $D^0$ and $J/\psi$ data
point to the same magnitude of \RgPb\ and their inclusion in the EPPS16 fit 
would likely result in a considerable reduction of its gluon uncertainty
by a factor as large as $1.7$, see \cf{diagram-f}.
For nCTEQ15, the effect seems less spectacular  but we should 
recall that the original nCTEQ15 values at $x$ below 
$10^{-3}$ are pure extrapolations. The dashed red lines in \cf{diagram-e} illustrates
this by showing two equally good fits~\cite{Stavreva:2010mw}, which are now 
excluded by the LHC HF \pPb\ data. Overall, 
the nCTEQ15 extrapolation to small $x$ is unexpectedly well confirmed 
by the charm(onium) data. 

Beside the mere observations of the nPDF-uncertainty reduction, our results have
two important physics interpretations. First, the LHC \pPb\ HF data give 
us the first real observation of gluon shadowing at small $x$ with $R_g^{\text{Pb}}$ 
smaller than unity --the no-shadowing null-hypothesis-- 
by more than 11.7 (10.9) and 7.3 (7.1) $\sigma$ at $x=10^{-5}$ and $\mu_F=2$~GeV for 
nCTEQ15 and EPPS16 using $D^0$ ($J/\psi$) data [see Figs.~\cf{diagram-e}-\ref{diagram-f}, left panels].
Our results thus quantitatively confirm the qualitative observations 
of~\cite{Abelev:2012ba,Guzey:2013xba,Khachatryan:2016qhq} indirectly made from $J/\psi$ photoproduction
on lead, which strictly speaking is sensitive to generalized parton distributions --not nPDFs--
and suffers from significant scale uncertainties~\cite{Ivanov:2004vd,Jones:2015nna}.
Second, our analysis corroborates the existence of
a gluon antishadowing~\cite{Frankfurt:1990xz}: $\RgPb>1$ for $x \simeq 0.1$.
This can be seen in Figs.\ \ref{diagram-e} and \ref{diagram-f} where the error band after reweighting
is smaller and more clearly separated from unity.
The analyzed LHC heavy quark(onium) data cover the $x$ region $7 \times 10^{-6} \lesssim x \lesssim 0.1$.
It is an interesting question how much of the antishadowing can be explained by direct data constraints
in the region $x \lesssim 0.1$ and how much of the effect is indirectly driven by the momentum sum rule correlating
a strong suppression at small $x$ with an enhancement in the antishadowing region.
We leave this question open for a future publication.

Finally, we consider the global coherence of the HF constraints
with other data (to be) included in nPDF global fits. We do it with nCTEQ15
of which 2 of us are authors. We thus have all the data at hand. 
First, let us observe that the agreement with the DIS NMC data~\cite{Arneodo:1996ru}, the only DIS set with a mild 
sensitivity to the gluon distribution, is not degraded in a statistically significant way.
The original $\chi^2/N_{\rm data}$, 0.58, becomes
(0.81, 0.58, 0.57) for $D^0$ with $\xi=(0.5,1,2)$, and is similar for other hadrons.
Clearly, the inclusion of HF data does not create any tension with the DIS data. 
One can also make a similar 
comparison for the $W/Z$ \pPb\ LHC data
whose impact on nCTEQ15 was recently studied~\cite{Kusina:2016fxy}.
The $\chi^2/N_{\rm data}$ of these data was found to be 2.43,
and after our HF reweighting it becomes
(2.14, 2.49, 3.11) for $D^0$.
With the same caveats as above, our reweighted nPDFs do not change the theory-data compatibility with the LHC $W/Z$ data.
The $\chi^2/N_{\rm data}$ of the $J/\psi$ PHENIX \RdAu\ results~\cite{Adare:2010fn,Adare:2012qf} with nCTEQ15 is (3.58, 2.55, 3.12) and
after our $J/\psi$ reweighting becomes
(1.81, 2.38, 2.77). 
This confirms the global coherence of the HF constraints.
Tables of these $\chi^2$ values
can be found as Supplemental Material.

\textit{Conclusion} -- 
In this Letter, we used, for the first time, experimental data for the 
inclusive HF [$D^0$, $J/\psi$, $B\rightarrow J/\psi$, $\Upsilon(1S)$]
production in \pPb\ collisions at the LHC to improve our knowledge of the gluon density inside heavy nuclei.
We compared the data with computations obtained in the
standard LT factorization framework endowed with
the two most recent globally fit nPDFs (nCTEQ15, EPPS16).
No other nuclear effects were included which are supposed to be of HT origin
and hence suppressed as inverse powers of the hard scale.
We found a good description of the LHC data with both nCTEQ15 and
EPPS16 nPDFs validating our theoretical framework.

By performing a Bayesian-reweighting analysis and studying 
the scale uncertainties, we demonstrated that the existing heavy quark(onium) data can 
significantly --and coherently-- reduce the uncertainty of the gluon density down 
to $x\simeq 7\times 10^{-6}$. For charm(onium), the gluons are shadowed 
with a statistical significance beyond 7~$\sigma$ at $\mu_F=2$ GeV and $x=10^{-5}$.
These data should thus be included in the next generation of global nPDF analyses.
While our results cannot rule out that other HT CNM effects were effectively 
``absorbed" into seemingly universal LT nPDFs, 
the observed consistent description of both the $D^0$ and $J/\psi$ 
data is far nontrivial since they may interact differently with the nuclear matter.

\acknowledgments
We are grateful to H.\ Paukkunen for the comments on using EPPS16 grids, to M.\ Cacciari for the private FONLL code, and to F.\ Arleo, E.G.\ Ferreiro, and P.\ Zurita for useful discussions.
The work is supported by the ILP Labex (ANR-11-IDEX-0004-02, ANR-10-LABX-63) and the COPIN-IN2P3 Agreement.

\bibliographystyle{utphys}

\bibliography{paper.bib}

\onecolumngrid

\appendix

\section{Supplemental material}

\subsection{Technicalities of data-driven approach}

The data-driven approach used here was first introduced in Ref.~\cite{Lansberg:2016deg}. Especially, the partonic amplitude is parameterized following Eq.(1) of \cite{Lansberg:2016deg}.  In principle, four parameters need to be determined from the proton-proton experimental data after convolving with the PDFs. The fit is done data set by data set for each scale choice. We summarized the fitted values of these parameters, $\kappa,\lambda,\langle P_T \rangle$, in Tab.~\ref{tabppfit} for the 4 particles ($D^0,J/\psi,B\rightarrow J/\psi, \Upsilon(1S)$) and 3 scale variations ($\mu_F=\xi \mu_0,\xi=0.5,1.0,2.0$) which we considered. We left the 4th parameter $n$ fixed to 2. 

\begin{table*}[!hbt]
\renewcommand{\arraystretch}{1.4}
\setlength{\tabcolsep}{12pt}
\begin{tabular}{|cc|cccc|}
\hline
& & $D^0$ & $J/\psi$ & $B\rightarrow J/\psi$ & $\Upsilon(1S)$ \\
\hline\hline
\multirow{3}{*}{$\kappa$} & $\xi=0.5$ & $1.26$ & $0.92$ & $0.25$ & $0.69$\\
& $\xi=1.0$ & $0.66$ & $0.56$ & $0.15$ & $0.77$\\
& $\xi=2.0$ & $0.50$ & $0.41$ & $0.13$ & $0.69$\\\hline
\multirow{3}{*}{$\lambda$} & $\xi=0.5$ & $2.97$ & $0.58$ & $0.09$ & $0.10$\\
& $\xi=1.0$ & $1.78$ & $0.30$ & $0.09$ & $0.08$\\
& $\xi=2.0$ & $1.39$ & $0.22$ & $0.08$ & $0.07$\\\hline
\multirow{3}{*}{$\langle P_T \rangle$} & $\xi=0.5$ & $0.01$ & $4.5$ (fixed) & $0.02$ & $13.5$ (fixed)\\
& $\xi=1.0$ & $0.09$ & $4.5$ (fixed) & $3.84$ & $13.5$ (fixed)\\
& $\xi=2.0$ & $0.03$ & $4.5$ (fixed) & $0.11$ & $13.5$ (fixed)\\
\hline
\end{tabular}
\caption{A summary of the fitted parameters $\kappa,\lambda,\langle P_T \rangle$ of our data-driven method.\label{tabppfit}}
\end{table*}

\subsection{Validation with FONLL}\label{validate}

\begin{figure}[!thb]
\begin{center}
\subfloat[nCTEQ15 nPDF]{\includegraphics[width=0.49\textwidth,draft=false]{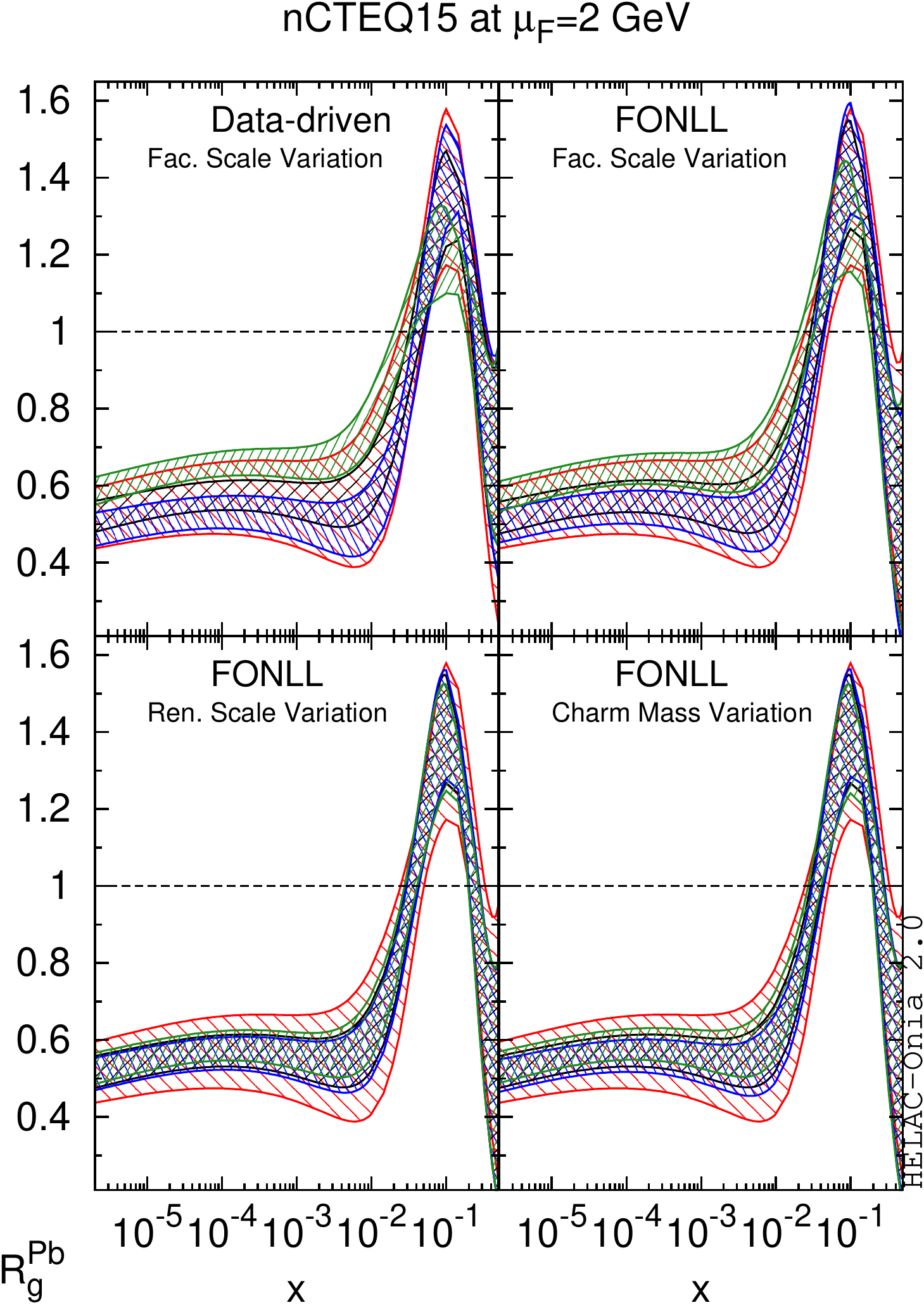}\label{fig:CompareFONLLa}}
\subfloat[EPPS16 nPDF]{\includegraphics[width=0.49\textwidth,draft=false]{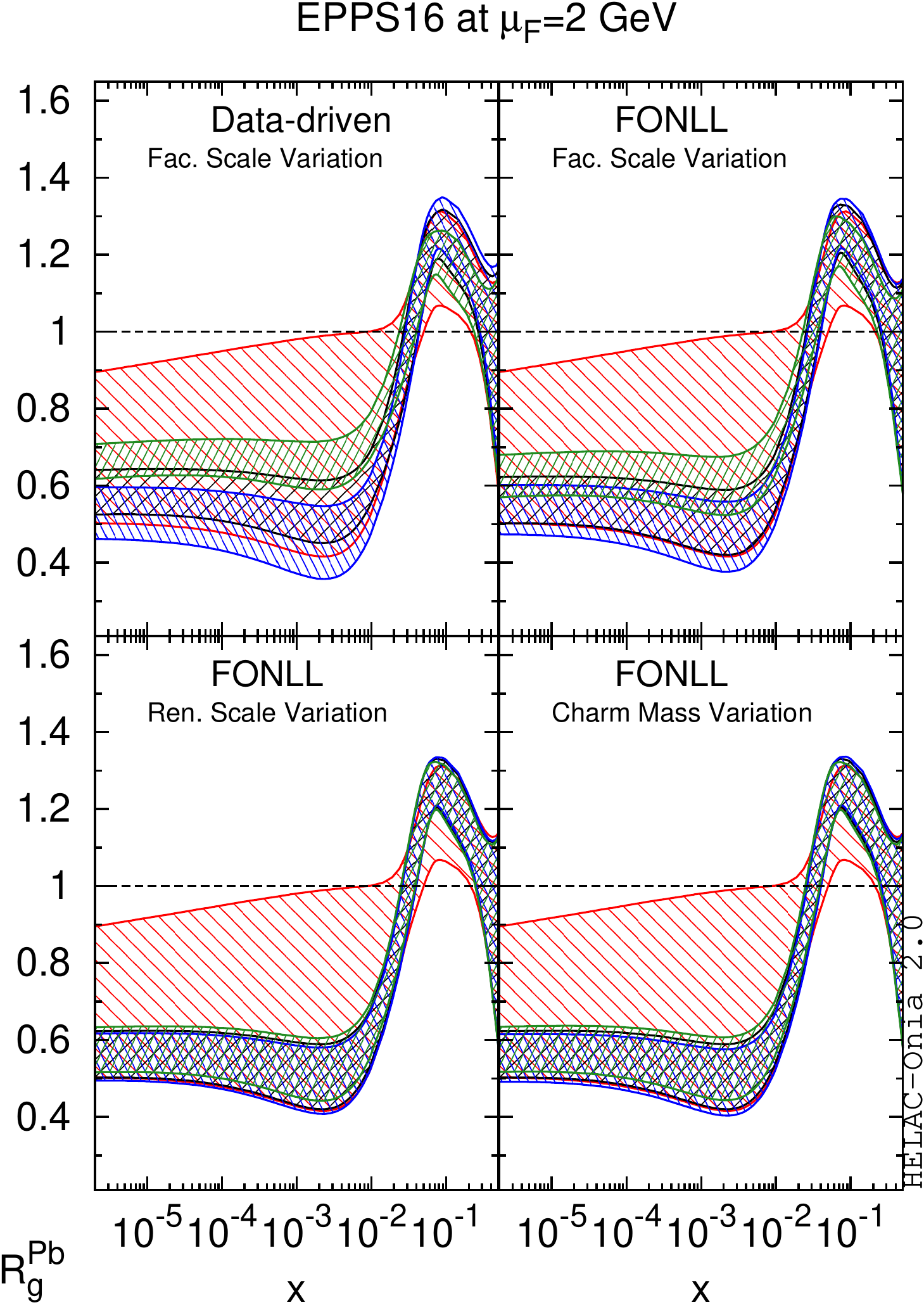}\label{fig:CompareFONLLb}}
\caption{Comparison of the reweighted nPDFs with $D^0$ data between our data-driven approach and FONLL with various theoretical uncertainties. The error bands due to nPDF uncertainty are given at $68\%$ CL level.
}
\label{fig:CompareFONLL}
\end{center}\vspace*{-1cm}
\end{figure}

We have explicitly checked the reweighted results from our data-driven approach with those from the FONLL perturbative calculation~\cite{Cacciari:1998it,Cacciari:2012ny} for open heavy-flavour production. The reweighted nPDFs from both approaches are shown in Fig.~\ref{fig:CompareFONLL} for the case of LHC $D^0$ data. In the data-driven approach, the only theoretical uncertainty to be considered is from the factorisation scale variation (shown in the top-left insets of Figs.~\ref{fig:CompareFONLLa} and \ref{fig:CompareFONLLb}). On the other hand, in the case of FONLL, besides the factorization scale uncertainty (shown in the top-right insets of Figs.~\ref{fig:CompareFONLLa} and \ref{fig:CompareFONLLb}), we also provide other theoretical uncertainties from the renormalisation scale (shown in the bottom-left insets of Figs.~\ref{fig:CompareFONLLa} and \ref{fig:CompareFONLLb}) and charm quark mass variations (shown in the bottom-right insets of Figs.~\ref{fig:CompareFONLLa} and \ref{fig:CompareFONLLb}). It confirms that for observables like $R_{p\rm{Pb}}$, the dominant theoretical uncertainty is from the factorisation scale variation, while the renormalisation scale and the charm mass uncertainties are largely cancelled. One also clearly sees that the results obtained using our data-driven method and FONLL are very similar confirming the validity of the data-driven method. The same situation holds also for the $B\rightarrow J/\psi$ case.

\subsection{$\chi^2$ numbers}

The $\chi^2$ values before and after reweighting are displayed in Tab.~\ref{tabchi2} for $\{D^0,J/\psi,B\rightarrow J/\psi, \Upsilon(1S)\}$ production in $p{\rm Pb}$ collisions at the LHC, together with total number of data points $N_{\rm data}$. As it is customary, these $\chi^2$ values do not account for any theoretical uncertainties. Regardless of the scale choice (i.e. $\xi$), $\chi^2/N_{\rm data}$ are around $1$ after reweighting, while $\chi^2/N_{\rm data}$ varies significantly with the original nPDFs. It is normal as our replicas are matching the $R_{p{\rm Pb}}$ vs $P_T,y$ data. We take the inclusive $J/\psi$ PHENIX $R_{dAu}$ results as a postdiction, where the $\chi^2$ numbers before and after reweighting are shown in Tab.~\ref{tabchi2others}. The compatibility between the theoretical calculations and the PHENIX data is further improved with the reweighted nPDFs. We also have checked the global coherence of the HF constraints with the LHC W/Z and DIS NMC data. The corresponding $\chi^2$ values are also shown in Tab.~\ref{tabchi2others}. No degradation is observed as the $\chi^2/N_{\rm data}$ values similar before and after the reweighting.

\begin{table*}[!t]
\renewcommand{\arraystretch}{1.4}
\setlength{\tabcolsep}{12pt}
\begin{tabular}{|cc|cccc|}
\hline
& & $D^0$ & $J/\psi$ & $B\rightarrow J/\psi$ & $\Upsilon(1S)$ \\
\hline\hline
& $N_{\rm data}$ & 38 & 71 & 37 & 12 \\\hline\hline
\multirow{3}{*}{{\rm Original nCTEQ15}} & $\xi=0.5$ & $142$ & $131$ & $39$ & $14$\\
& $\xi=1.0$ & $39$ & $63$ & $23$ & $11$\\
& $\xi=2.0$ & $63$ & $90$ & $15$ & $11$\\\hline
\multirow{3}{*}{{\rm Reweighted nCTEQ15}} & $\xi=0.5$ & $56$ & $46$ & $14$ & $13$\\
& $\xi=1.0$ & $56$ & $53$ & $11$ & $11$\\
& $\xi=2.0$ & $56$ & $46$ & $9$ & $11$\\\hline\hline
\multirow{3}{*}{{\rm Original EPPS16}} & $\xi=0.5$ & $53$ & $62$ & $9$ & $10$\\
& $\xi=1.0$ & $140$ & $150$ & $7$ & $10$\\
& $\xi=2.0$ & $218$ & $220$ & $8$ & $11$\\
\hline
\multirow{3}{*}{{\rm Reweighted EPPS16}} & $\xi=0.5$ & $37$ & $59$ & $7$ & $10$\\
& $\xi=1.0$ & $37$ & $59$ & $7$ & $10$\\
& $\xi=2.0$ & $37$ & $59$ & $7$ & $11$\\
\hline
\end{tabular}
\caption{Total numbers of $R_{p{\rm Pb}}$ vs $P_T,y$ data points used for reweighting and the corresponding $\chi^2$ values before and after reweighting. No theoretical uncertainties are taken into account when evaluating $\chi^2$.\label{tabchi2}}
\end{table*}

\begin{table*}[!t]
\renewcommand{\arraystretch}{1.4}
\setlength{\tabcolsep}{12pt}
\begin{tabular}{|c|cc|c|cccc|}
\hline
& & & \multirow{2}{*}{{\rm Original}} & \multicolumn{4}{c|}{Reweighted}\\\cline{5-8}
& & & & $D^0$ & $J/\psi$ & $B\rightarrow J/\psi$ & $\Upsilon(1S)$ \\
\hline\hline
\multirow{6}{*}{{\rm PHENIX} $J/\psi$ ($N_{\rm data}=74$)} & \multirow{3}{*}{{\rm nCTEQ15}} & $\xi=0.5$ & $265$ & $-$ & $134$ & $-$ & $-$\\
& & $\xi=1.0$ & $189$ & $-$ & $176$ & $-$ & $-$\\
& & $\xi=2.0$ & $231$ & $-$ & $205$ & $-$ & $-$\\\cline{2-8}
& \multirow{3}{*}{{\rm EPPS16}} & $\xi=0.5$ & $133$ & $-$ & $138$ & $-$ & $-$\\
& & $\xi=1.0$ & $207$ & $-$ & $167$ & $-$ & $-$\\
& & $\xi=2.0$ & $263$ & $-$ & $209$ & $-$ & $-$\\
\hline
\multirow{3}{*}{{\rm LHC} W/Z ($N_{\rm data}=102$)} & \multirow{3}{*}{{\rm nCTEQ15}} & $\xi=0.5$ & \multirow{3}{*}{$248$} & $218$ & $230$ & $212$ & $229$\\
& & $\xi=1.0$ & & $254$ & $271$ & $214$ & $238$\\
& & $\xi=2.0$ & & $317$ & $332$ & $219$ & $243$\\\hline
\multirow{3}{*}{{\rm NMC} $F_2^{\rm Sn}/F_2^C$ ($N_{\rm data}=111$)} & \multirow{3}{*}{{\rm nCTEQ15}} & $\xi=0.5$ & \multirow{3}{*}{$65$} & $93$ & $98$ & $86$ & $70$\\
& & $\xi=1.0$ & & $65$ & $66$ & $78$ & $67$\\
& & $\xi=2.0$ & & $62$ & $62$ & $71$ & $65$\\\hline
\multirow{3}{*}{{\rm NMC} $F_2^{\rm Pb}/F_2^C$ ($N_{\rm data}=14$)} & \multirow{3}{*}{{\rm nCTEQ15}} & $\xi=0.5$ & \multirow{3}{*}{$8$} & $8$ & $8$ & $8$ & $7$\\
& & $\xi=1.0$ & & $7$ & $6$ & $7$ & $7$\\
& & $\xi=2.0$ & & $9$ & $8$ & $7$ & $8$\\\hline
\end{tabular}
\caption{Comparison of $\chi^2$ values for inclusive $J/\psi$ $d{\rm Au}$ PHENIX data, $W/Z$ $p{\rm Pb}$ LHC data and NMC data before and after reweighting. No theoretical uncertainties are taken into account for evaluating $\chi^2$.\label{tabchi2others}}
\end{table*}

\subsection{Effective number of replicas}

The reliability of the reweighting procedure can be estimated by the effective number of replicas $N_{\rm eff}$ after reweighting (see Eq.(14) in Ref.~\cite{Kusina:2016fxy}). It provides an estimation of the number of replicas effectively contributing to the reweighting procedure. If $N_{\rm eff}/N_{\rm rep}\ll 1$ with $N_{\rm rep}$ the number of original replicas, the reweighting procedure becomes inefficient and a new global fit is necessary. We provide the values of $N_{\rm eff}$ in Tab.~\ref{tabNeff} based on our original $N_{\rm rep}=10^4$ replicas. Among the $24$ reweighting results (2 nPDFs, 4 data sets and 3 factorisation scale choices $\mu_F=\xi \mu_0$), we conclude that we always have $N_{\rm eff} > 3000$, which confirms the reliability of our reweighting results.

\begin{table*}[!t]
\renewcommand{\arraystretch}{1.4}
\setlength{\tabcolsep}{12pt}
\begin{tabular}{|cc|cccc|}
\hline
& & $D^0$ & $J/\psi$ & $B\rightarrow J/\psi$ & $\Upsilon(1S)$ \\
\hline\hline
\multirow{3}{*}{{\rm nCTEQ15}} & $\xi=0.5$ & $3063$ & $3423$ & $6584$ & $9508$\\
& $\xi=1.0$ & $5573$ & $5906$ & $7859$ & $9830$\\
& $\xi=2.0$ & $5353$ & $5479$ & $8625$ & $9929$\\\hline
\multirow{3}{*}{{\rm EPPS16}} & $\xi=0.5$ & $3116$ & $3304$ & $7914$ & $9724$\\
& $\xi=1.0$ & $3979$ & $4204$ & $8444$ & $9875$\\
& $\xi=2.0$ & $4226$ & $4462$ & $8783$ & $9932$\\
\hline
\end{tabular}
\caption{Summary of $N_{\rm eff}$ after performing reweighting with $10^4$ original replicas.\label{tabNeff}}
\end{table*}

\end{document}